\documentclass[]{aastex63}

\usepackage{mathrsfs}
\usepackage{amsmath}
\usepackage{booktabs}
\usepackage{lineno}
\ifdefined\DeclareUnicodeCharacter
  \DeclareUnicodeCharacter{FB01}{\nobreakspace}
\fi
\ifdefined\DeclareUnicodeCharacter
  \DeclareUnicodeCharacter{FB02}{\nobreakspace}
\fi
\newcommand{\mafont}{
  \color{red}
}
\DeclareTextFontCommand{\ma}{\mafont}
\usepackage[normalem]{ulem}

\newcommand\maout{\bgroup\markoverwith
{\textcolor{red}{\rule[.5ex]{2pt}{0.4pt}}}\ULon}

\received{\today}
\revised{}
\accepted{}

\shorttitle{Runaway return current}
\shortauthors{Alaoui, Holman, Swisdak}

\graphicspath{{./}{figures/}}

\begin{document}

\title{Reduction of the downward energy flux of non-thermal electrons in the solar flare corona due to co-spatial return current losses}

\correspondingauthor{Meriem Alaoui}
\email{alaoui@umd.edu}

\author[0000-0003-2932-3623]{Meriem Alaoui}
\affiliation{Institute for Research in Electronics and Applied Physics\\
8279 Paint Branch Dr, College Park, MD 20740}
\affiliation{NASA Goddard Space Flight Center\\8800 Greenbelt Rd, Greenbelt, MD 20771}

\author{Gordon D. Holman}
\affiliation{Laboratory for Solar Physics (Emeritus)\\NASA Goddard Space Flight Center\\8800 Greenbelt Rd, Greenbelt, MD 20771}
\author{M. Swisdak}
\affiliation{Institute for Research in Electronics and Applied Physics\\
8279 Paint Branch Dr, College Park, MD 20740}

\begin{abstract}
High energy electrons carry much of a solar flare's energy. Therefore, understanding changes in electron beam distributions during their propagation is crucial. A key focus of this paper is how the co-spatial return current reduces the energy flux carried by these accelerated electrons. We systematically compute this reduction for various beam and plasma parameters relevant to solar flares. Our 1D model accounts for collisions between beam and plasma electrons, return current electric-field deceleration, thermalization in a warm target approximation, and runaway electron contributions. The results focus on the classical (Spitzer) regime, offering a valuable benchmark for energy flux reduction and its extent. Return current losses are only negligible for the lowest nonthermal fluxes. We calculate the conditions for return current losses to become significant and estimate the extent of the modification to the beam's energy flux density. We also calculate two additional conditions which occur for higher injected fluxes: (1) where runaway electrons become significant, and (2) where current-driven instabilities might become significant, requiring a model that self-consistently accounts for them. Condition (2) is relaxed and the energy flux losses are reduced in the presence of runaway electrons. All results are dependent on beam and co-spatial plasma parameters. We also examine the importance of the reflection of beam electrons by the return-current electric field.  We show that the interpretation of a number of flares needs to be reviewed to account for the effects of return currents.
\end{abstract}

\section{Introduction} \label{sec:intro}
Understanding the physical mechanisms responsible for the explosive release of energy in solar flares is a critical issue in heliophysics. Flares rapidly and efficiently accelerate large numbers of electrons to energies in the tens to hundreds of keV \citep[e.g.,][]{2019SoPh..294..105A}, with total energies reaching $\sim$10$^{33}$ erg \citep{2012ApJ...759...71E} and $\sim$10$^{37}$ erg \citep{2020AJ....159...60G,2024LRSP...21....1K} in solar and stellar flares, respectively. A significant fraction of this energy is carried by nonthermal electrons, as evidenced by their X-ray radiation. As these electrons propagate through the atmosphere they collide with electrons and ions in the ambient plasma, resulting in plasma heating, fast hydrodynamic flows, and bursts of hard X-ray (HXR) radiation. The electron beam constitutes a very large electrical current that is neutralized by a co-spatial return current (RC) \citep[e.g.,][]{1977A&A....55...23H,1977ApJ...218..306K,2006ApJ...651..553Z,2008A&A...486..325K,2012ApJ...745...52H,2017ApJ...851...78A}, which can significantly alter the nonthermal beam’s propagation through its interaction with the finite resistivity of the co-spatial atmosphere. 

In addition, atmospheric response models also depend on an accurate characterization of the energy injected into the chromosphere, as well as the initial temperature and density profiles of a plasma flux tube.  
To effectively interpret HXR emission and recover the accelerated electron energy distribution and the atmospheric response, it is crucial to have a comprehensive understanding of the mechanisms governing the propagation of electron beams. Unfortunately, these mechanisms are frequently simplified to such an extent that the energy distribution and overall energy content of the accelerated electrons can be significantly misrepresented.

Other effects might also be energetically important enough to modify the energy distribution along the loop legs and therefore the energy deposition, the magnitude of the return current electric field, and the deduced accelerated electron distribution. For example, MHD turbulence can enhance the local electrical resistivity along the legs of the loop \citep[][]{2016ApJ...824...78B,2018ApJ...862..158E,2018ApJ...865...67E}. \citet[][]{2022ApJ...931...60A} included the modification of the \emph{thermal} conduction by this mechanism and found that a model with a small enhancement to the thermal resistance by turbulent scattering best explained the Doppler profiles for an X-class flare. Other examples include: magnetic mirroring, which can reduce the number of downward propagating electron in the beam and decrease the mean free path of electrons \citep[e.g.,][]{2014A&A...563A..51V,2016ApJ...832...63D}; wave-particle interactions, which might be important in some cases under specific conditions of the co-spatial plasma \citep[][]{2008A&A...487..337B,2017ApJ...851...78A}; and modification of the beam distribution by interactions with Langmuir waves \citep[e.g.,][]{1987ApJ...321..721H,2012A&A...544A.148K,2013A&A...550A..51H}. In \citet[][hereafter AH17]{2017ApJ...851...78A}, it was shown that return current-driven instabilities are unlikely to be responsible for the HXR spectral break in the majority of flares. A notable exceptions is SOL2006-12-06T18:30, an X-class flare that will be discussed further in section~\ref{sec:obs}. Those results also need to be reviewed in light of our new understanding of the beam/RC dynamics.

Importantly, any significant mechanism will also affect the beam/RC system and thereby the energy losses along the direct beam's path from injection to its thermalization. Significant progress has been achieved in understanding the beam/RC propagation in flares, although this understanding is still rarely used in the interpretation of solar observations. 
 
Using the beam/return current propagation model in \citet[][ hereafter, A21]{2021ApJ...917...74A}, we will show the conditions of the nonthermal electron beam and co-spatial plasma in which it propagates (the return current plasma) where the return current substantially affects the energy of the nonthermal electron beam. 

The model is 1D and includes both Coulomb collisions between the beam and co-spatial plasma electrons as well as return current deceleration of the beam electrons through the RC electric field, including when this electric field is large enough to accelerate runaway electrons out of the ambient plasma.  The simulation code solves the equations describing the energy losses of an electron as it propagates down the legs of a flaring magnetic flux tube as part of an initial power-law distribution. Electrons are thermalized when their energy reaches a few times the thermal energy. This ensures that we account for the loss of electrons from the direct downward propagating beam as they get thermalized. Importantly, the model accounts for transitions between fully collisional and quasi-collisionless regimes of the beam/RC propagation, where the resistivity is treated with the Spitzer-Braginskii formulation, and the number of runaway electrons accelerated at any position along the loop legs is small to avoid distorting the RC plasma from a Maxwellian distribution plus a nonthermal tail. 

For the lowest electron flux densities, the RC deceleration is insufficient to affect the beam's energy from its injection at the looptop to the chromosphere. This will be quantified for a large range of co-spatial parameters and electron distributions.

Knowledge of the energy deposition profile is a key element in determining the response of the solar atmosphere to the injection of an electron beam  \citep{bradshaw2013,reep2019,2015ApJ...809..104A} and interpretation of observations of the chromospheric emission \citep[e.g.,][for recent reviews of energy transport at the chromosphere]{2024arXiv240413214K,2022FrASS...960856K,2023FrASS...960862K}.

 For this paper, we will address the following questions: \begin{enumerate}
 \item{How much of the nonthermal energy flux density is lost in the coronal legs of the loop before the electrons reach the flare footpoints?} 
 \item{What fraction of the electrons returning to the acceleration region are suprathermal, i.e., runaways accelerated by the return-current electric field?} 
 \item{Under which conditions can the return current propagation be explained by a collisional or a quasi-collisionless regime, where collisionless RC propagation includes runaway electron acceleration and/or RC driven instabilities?}
 \item{Can the temperature and density along the legs of a flaring loop be constrained by accounting for return current effects?}
 \item{Is a significant fraction of the beam electrons mirrored by the return-current electric field?}
 \end{enumerate}
 
The goal is to determine the conditions in solar flares for which the beam/return current system affects the beam energetics, and to understand how this occurs. This provides a reference for when return currents significantly reduce the energy flux in the corona and therefore affect the energy flux density deposited below the transition region. This will improve our prediction of the atmospheric response by using the beam/RC system propagation model as input into the widely used radiative hydrodynamic codes and help predict the accelerated electron distribution.

We model the electron transport in a flaring loop with an electron plasma temperature and density ranging from 1~MK to 33~MK, and $6\times10^9$ to $3\times10^{10}$ cm$^{-3}$, respectively. From the detailed particle transport calculation using a range of electron beam parameters, we compute the corresponding energy flux density at the footpoints using the model in A21 which accounts for return current losses in the presence of runaway electrons accelerated out of the co-spatial plasma, warm target effects and Coulomb collisions between electrons in the injected nonthermal beam electrons and ambient (co-spatial return current) plasma. We show these results in section~\ref{sec:main}. These results should be used to determine whether the return current is significant in modifying the energy flux density and the extent of this modification, as well as the significant propagation mechanisms. 
In other words, when are return current losses and runaway electrons significant.
 
In section~\ref{sec:mirror}, we derive and discuss upper limits to the reduction of downward propagating electrons and their associated energy flux density by electric field mirroring of beam electrons. This mirroring does not occur in our 1D model and, therefore, is not taken into account. 
 
In Section~\ref{sec:instability}, we discuss the thresholds for generating current-driven instabilities and show that the transition from collisional to quasi-collisionless regimes involves the acceleration of runaway electrons for $T_e\lesssim 4T_i$, where $T_e$ and $T_i$ are the co-spatial (ambient) electron and ion temperatures. In addition, the thresholds for generating the current-driven instabilities are increased in the presence of runaways. The role of return currents in generating Langmuir waves will be discussed in an upcoming paper. 
 
In section~\ref{sec:obs}, we discuss the relevance of our calculations for various previously published X-class flares with reported large injected flux densities flares. A discussion and implications for future work regarding the acceleration region and the atmospheric response to the injection of an electron beam are covered in section~\ref{sec:discussion}.

\section{Fraction of energy flux density injected at looptop that reaches loop footpoints}\label{sec:main}

\subsection{Beam/return current propagation model}
We are using the model in \citet[][]{2021ApJ...917...74A} which self-consistently solves the energy loss equation of an electron beam propagating in a 1D plasma under the influence of Coulomb collisions and return current deceleration, including when the return current electric field is large enough to freely accelerate runaway electrons out of the ambient plasma (co-spatial return-current plasma). The model is steady-state which means that the beam/return current system had enough time to balance the beam and return currents everywhere along the loop legs. This happens on time scales on the order of the electron-ion collision time \citep[][]{1990A&A...234..496V,2009A&A...504.1057S}. The steady state solution is appropriate as long as the time scale of beam injection is long enough compared to the collisional time scale.

We also account for warm target effects using the warm target approximation: a target is considered "cold" if the energy in the beam electrons is much higher than the ambient plasma with which they interact. If electrons lose energy through RC losses, for example, the corona can become a "warm" target to the lower energy electrons. The beam electrons whose energy becomes close to that of the co-spatial plasma are lost from the nonthermal beam, which reduces the total number flux density. We will show throughout the paper the importance of including this effect. The thermalization distance $x_{RC}$ is where the electrons with initial energy equal to the low-energy cutoff lose enough of their energy to become equal to $E_{th} = \delta k_BT$ with $\delta$ the spectral index of the nonthermal electron energy distribution and $T$ the ambient plasma temperature. This energy is where energy diffusion dominates over the dynamic friction force for a single power-law electron distribution, i.e., where the thermalization dominates over energy loss by collisions \citep[][]{2015ApJ...809...35K,2019ApJ...880..136J,2020ApJ...902...16A}.

The model can account for Coulomb collisions alone (\texttt{CC only}),
return currents alone without Coulomb collisions (\texttt{RC only}), or return currents and collisions without inclusion of runaway electrons (\texttt{RC+CC}), and finally the case which includes all effects (\texttt{RC+CC+Run}).\footnote{These cases were named differently in \cite[][]{2021ApJ...917...74A}.} All models include thermalization of beam electrons if their energy reaches $\delta kT$. 

Input parameters are the injected nonthermal distribution, assumed to be a single power-law with a sharp low-energy cutoff (injected nonthermal number flux density in electrons cm$^{-2}$ s$^{-1}$, electron spectral index $\delta$, and the low-energy cutoff E$_c$), and the co-spatial plasma temperature and density profiles. In the majority of the calculations we choose a constant temperature and density in the corona for simplicity, but we also show how a gradient in temperature and density affect the solutions in section~\ref{sec:example}.

We obtain a grid of electron energy evolution along the loop leg $E(E_0,x)$, where $E_0$ is the initial energy of an electron in the nonthermal beam, and the electron energy distribution as a function of space $F(E,x)$. We can then calculate various quantities like the nonthermal energy flux density $\mathscr{F}(x)$ and the volumetric power lost by the beam, which is equivalent to the volumetric heating rate in the absence of runaway electrons.
The energy flux density is calculated as follows: \begin{equation}\mathscr{F}(x)=\int_{E_c(x)}^{\infty} E'\, F(E',x)\, dE', \label{eq:eflux}\end{equation} where $F(E,x)$ is the electron energy distribution at position $x$, and the injected electron energy distribution $F_0(E_0)=F(E,x=0)=(\delta-1)\,E_c^{\delta-1}\,F_{e0}\,E_0^{-\delta}$, where $F_{e0}$ is the total electron flux density.

In the following sections we show the energy flux density at the transition region  (TR) for various beam and co-spatial parameters. We discuss how the solutions are affected by these parameters, the importance of warm target effects due to both the return current deceleration (beam energy losses due to the RC electric force) and Coulomb collisions, and how including the effect of accelerating runaway electrons changes where and how much energy is deposited in the solar or stellar atmospheres. First, an example is discussed in detail (section~\ref{sec:example}), where we show the spatial evolution of the volumetric power lost by the beam, the heating rates and the energy flux density. This is useful for quantifying the heating rate, which can be used as input into radiative hydrodynamic codes \citep[e.g.,][]{1992ApJ...397L..59C,2015ApJ...809..104A,2013ApJ...778...76R}, and the extent of the error made when neglecting return currents with and without runaways. Then we focus on integrated quantities like the potential drop at the transition region and the runaway fraction at the looptop, or quantities at specific locations in the loop such as the flux density at the transition region, and the maximum RC electric field, in section~\ref{sec:systematic}. This is useful to systematically show thresholds for where return current losses are non-negligible and where runaway electrons become significant as a function of the wide range of input parameters relevant to solar and stellar flares.  

\subsection{Example}\label{sec:example}
\begin{figure*}[bth!]
\centering
  \includegraphics[width=0.98\textwidth]{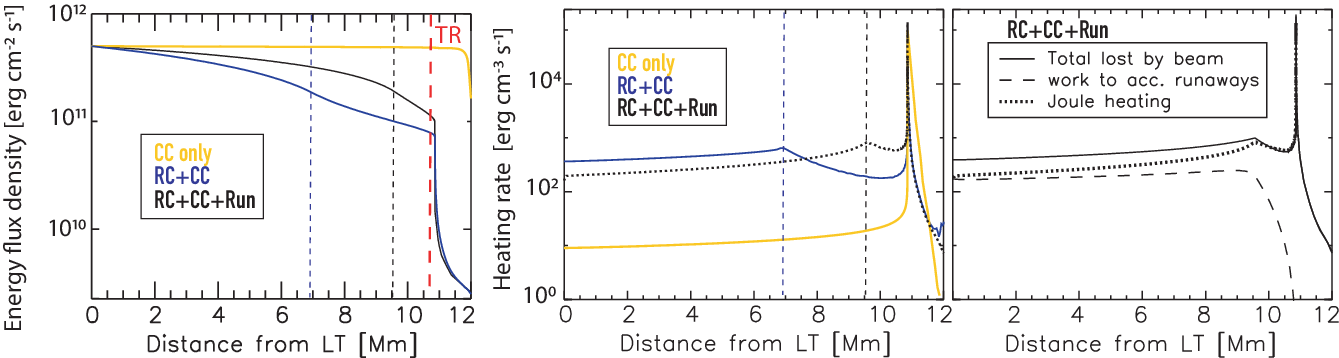}
   \caption{{\textit{Left:}} Energy flux density as a function of distance along the loop leg, for the same injected electron beam and atmosphere, using only \texttt{CC} (yellow), \texttt{RC+CC} without including runaway electrons (blue), and the full solution \texttt{RC+CC+Run} (black). The red vertical line shows the energy flux density at the transition region (TR) which will be used in the remainder of the paper. The thermalization distance $x_{RC}$ using \texttt{RC+CC} and \texttt{RC+CC+Run} is represented by the blue and black vertical lines, respectively. The rate of energy flux density reduction is slowed below $x_{RC}$, as a result of number flux density loss from the injected electron beam. {\textit{Middle:}} The corresponding volumetric power (heating rates) for all three solutions.  {\textit{Right:}} Volumetric power in the case \texttt{RC+CC+Run}. The solid curve is the total power lost by the beam, which is further divided into two components, the power used to accelerate the runaway electrons (dotted) and Joule heating (dashed).}
  \label{fig:heatrate}
\end{figure*} 
We compare the energy flux density deposition and the volumetric power along the legs of a half loop using three different models for energy losses: The first scenario, termed \texttt{CC only}, exclusively considers Coulomb collisions between the beam and background electrons. In the second scenario, denoted as \texttt{RC+CC}, the calculation includes self-consistent return currents and collisions, excluding the influence of runaway electrons accelerated out of the ambient plasma. Finally, the third scenario, \texttt{RC+CC+Run}, accounts for the presence and impact of runaway electrons accelerated by the return current electric field out of the return-current co-spatial plasma. Warm target effects are included in all three cases. 

Figure~\ref{fig:heatrate} shows calculations using an injected electron number flux density $F_{e0}=8.3\times10^{18}$ cm$^{-2}$ s$^{-1}$, corresponding to an injected energy flux density\footnote{We will adopt the notation 5F11 to signify $5\times10^{11}$ erg cm$^{-2}$s$^{-1}$. This is sometimes used in the solar flare community in studies of the thermal response of a plasma to the injection of energy into the solar atmosphere, and has been used for both the total energy density erg cm$^{-2}$ or energy flux density erg cm$^{-2}$ s$^{-1}$.}  $\mathscr{F}_0=5\times10^{11}$ erg cm$^{-2}$s$^{-1}$, low-energy cutoff of 25~keV, and electron spectral index $\delta=4$, into a 12~Mm loop with apex temperature and density of 3~MK and 7.5$\times10^{9}$ cm$^{-3}$. The temperature and density profiles used here are the same as the cool temperature atmosphere in Figure 1 of \citet{2020ApJ...902...16A}.

The left panel shows $\mathscr{F}(x)$, the spatial evolution of the energy flux density as a function of distance from the injection  at the looptop (LT).

The energy flux deposition at the transition region (TR), represented by the red dashed line, is almost 100$\%$ of the injected energy flux density if only collisions are taken into account, while only 16$\%$ and 20$\%$ reach the chromosphere when return currents are taken into account without and with accounting for runaways, respectively. 

The blue and black dashed vertical lines show where the nonthermal beam starts to be thermalized in the \texttt{RC+CC} and \texttt{RC+CC+Run} models, respectively. 

\citet[][]{2021ApJ...917...74A} showed that neglecting runaway electrons results in overestimating the return current electric field necessary to balance the beam current density. Since the local electric field is lower in \texttt{RC+CC+Run}, the beam electrons start to be thermalized lower down the loop leg compared to the case where runaways are neglected, \texttt{RC+CC}.

Two crucial points emerge from these analyses: (1) both return current losses and runaway electrons are significant in this example; and (2) while the flux density that reaches the TR in both the \texttt{RC+CC} and \texttt{RC+CC+Run} cases differ by only $\sim4\%$, the details of where in the corona the energy is deposited remains an important consideration. 

This is evident in the middle and right panels. When runaway electrons are neglected, all the power lost by the beam heats the ambient plasma. This is represented by the yellow and blue curves in the middle panel. However, if runaway electrons are significant, i.e., if the normalized RC electric field is high enough to freely accelerate a significant runaway tail, then the power lost by the beam (solid black line in both panels) will both heat the ambient plasma through Joule heating (dotted curve) and Coulomb collisions as well as accelerate the runaway electrons (dashed curve in right panel). The normalized RC electric field is ${\mathscr{E}_{RC}}/{\mathscr{E}_{D}}$, where $\mathscr{E}_D = 4\pi e^3  \ln\Lambda {n_e}/({k_B T})$ is the Dreicer field (cf. equation 3 in A21). Although this volumetric power lost by the beam is calculated self-consistently, we can separate its various components as follows:
 \begin{equation} \begin{split}Q(x)=-\frac{d}{dx} \int_{E_c(x)}^\infty E\, F(E,x)\,dE\, \\ =\frac{\mathscr{E}_{RC}^2(x)}{\eta(x)}+\mathscr{E}_{RC}(x)\times J_{run}(x) +Q_c(x),\end{split}\label{eq:power}\end{equation} 
where $F(E,x)$ is the electron energy distribution at position $x$ in electrons cm$^{-2}$ s$^{-1}$keV$^{-1}$, $\mathscr{E}_{rc}$ is the return current electric field, $J_{run}$ the runaway current, $\eta$ the resistivity and $Q_c$ the heating rate by Coulomb collisions.
 
The right panel shows the components of the power lost by the beam, where the dashed and dotted curves are the Joule heating (first term on the right hand side of equation~\ref{eq:power}) and the power lost by the beam which goes into accelerating runaway electrons (middle term of the equation). The highest fraction of the nonthermal beam's volumetric power that serves to accelerate runaway electrons is at the looptop where the maximum flux of the RC is carried by runaway electrons.

\subsection{Systematic calculations}
\label{sec:systematic}
We repeat the previous analysis for a wide range of beam parameters and co-spatial plasma conditions. This is meant to show enough conditions to quantify the energy flux density reduction even under different conditions not presented in this paper, and understand the behavior of the beam and return current along their propagation. 

Because return-current losses are sensitive to plasma resistivity ($\eta\propto T^{-3/2}$), and the injected \emph{number} flux density (electrons cm$^{-2}$ s$^{-1}$), we will investigate those parameters at higher resolution. We choose to plot the results as a function of the \emph{energy} flux density, as it is the quantity most widely used by the flare community studying radiative hydrodynamics, although the number flux (electrons s$^{-1}$) is more easily deduced from HXR observations. Knowledge of the loop cross-sectional area is necessary to infer the beam's current density, which is balanced by the return current density at every position along the loop. Different panels show the dependence on the spectral index, loop length and co-spatial density. We will show the results for two spectral indices (4 and 7), two densities ($6\times10^{9}$cm$^{-3}$ and $3\times10^{10}$cm$^{-3}$) and three loop half-lengths. Contrary to Figure~\ref{fig:heatrate} where the atmosphere consists of a gradient temperature and density profiles in a 12~Mm loop half-length, the temperatures and densities in this section are taken to be constant in the corona and ramp down and up, respectively, below the transition region. This is done for simplicity since there are an infinite number of ways the temperature and density profiles evolve as energy is deposited along the legs of the loop.

\subsubsection{Energy flux density at the transition region}

\begin{figure}[bth!]
\centering
   \includegraphics[width=0.49\textwidth]{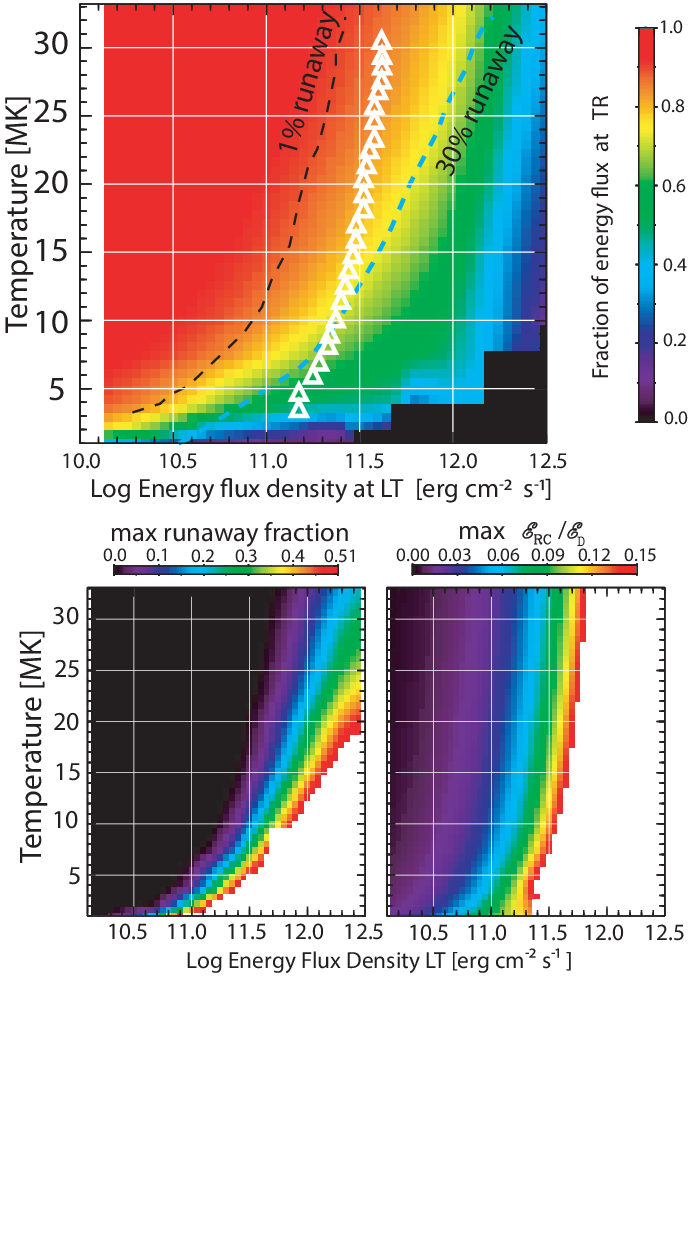} 
   \caption{Top: Fraction of energy flux density $\mathscr{F}_{TR}/\mathscr{F}_0$ as a function of injected energy flux density at the looptop $\mathscr{F}_0$ in erg cm$^{-2}$ s$^{-1}$ on the x-axis, the coronal temperature in MK on the y-axis. The dashed black and blue curves represent where the fraction of runaway electrons at the looptop (i.e., the maximum fraction of runaways) is 1$\%$ and 30$\%$, respectively. The white triangles show the curve of maximum RC electric field of 0.12$\mathscr{E}_{D}$.  Bottom left: Maximum fraction of the RC flux carried by runaways, Bottom right: Maximum normalized RC electric field.  }
  \label{fig:20keV_one}
\end{figure}

Figure~\ref{fig:20keV_one} shows the fraction of energy flux density which reaches the TR: ${\mathscr{F}_{TR}}/{\mathscr{F}_0}$. The parameters used are a 20 keV low-energy cutoff, $\delta=4$, loop half length 20~Mm and co-spatial density of $6\times10^9$ cm$^{-3}$. 

As expected, the largest reductions are at higher injected fluxes and lower co-spatial temperatures. The lower right black region is not well represented by our model, so the solutions were not calculated. In fact, for the remainder of the paper, we will systematically plot the normalized maximum return current electric field for which our model is no longer accurate because runaway electrons can dominate the co-spatial return current distribution.  The return current energy distribution is expected to differ significantly from a Maxwellian with an accelerated tail of runaways \citep[cf. Figure 1 and discussion in section 5.1 in][]{2021ApJ...917...74A}. The white triangles show where $\mathscr{E}_{RC}/\mathscr{E}_D=0.12$. The curves representing the fractions $1\%$ and $30\%$ of the return current flux carried by runaway electrons are plotted in the dashed black and blue curves, respectively. 
More details are represented in the lower panels of Figure~\ref{fig:20keV_one}, where the left is the maximum fraction of runaways and the right is the maximum normalized RC electric field.

At lower temperatures, a smaller increase in flux produces a larger runaway fraction. For example, it takes an order of magnitude increase in flux to increase the runaway fraction from 1\% to 30$\%$ at 25~MK but only a factor of 3 to do the same at 5~MK. This is because the RC electric field is higher in a plasma with a higher resistivity i.e., a cooler plasma.

\subsubsection{Potential drop at the transition region}
\label{sec:systematic2}

\begin{figure}[bth!]
\centering
   \includegraphics[width=0.495\textwidth]{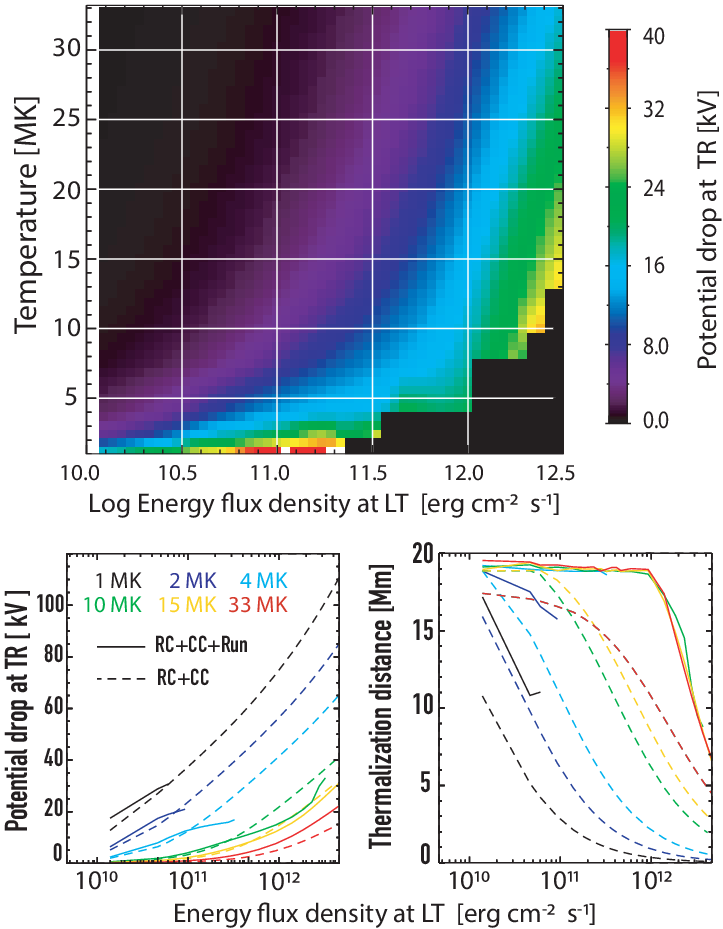}
   \caption{{\textit{top:}} Return current potential drop at the transition region (TR) as a function of injected energy flux density at the looptop $\mathscr{F}_0$ in erg cm$^{-2}$ s$^{-1}$ on the x-axis, and the coronal temperature in MK on the y-axis. This corresponds to the same parameters used in figure~\ref{fig:20keV_one}. \textit{Bottom:} The potential drop is comparable with and without runaways, however without runaways the beam is thermalized higher in the loop. In addition the energy deposition is overestimated (see text for explanation). {\textit{Bottom left:}} Potential drop at the TR as a function of injected flux density. The solid and dashed lines use \texttt{RC+CC+Run} and \texttt{RC+CC}, respectively. {\textit{Bottom right:}} Distance ($x_{RC}$) at which the beam starts being thermalized in Mm, this is where an electron with initial energy equal to the low-energy cutoff reaches $\delta\, kT$. The loop half length is 20~Mm.} 
  \label{fig:potential_one}
\end{figure}

Figure~\ref{fig:potential_one} shows the corresponding potential drop integrated from the LT to the TR in the full model. Note that the potential drop is also the maximum energy gained by runaway electrons at the looptop. For example, for injected fluxes of F11 and 2F11 into a plasma of initial temperature 7.5~MK, the maximum energy gained by runaway electrons is 6 and 9.5~keV, respectively.

The lower panels show the potential drop at the TR and the thermalization distance from the looptop on the left and right panels, respectively. The calculations use \texttt{RC+CC+Run} (solid curves) and \texttt{RC+CC} (dashed curves).

The total potential drop (at the TR) can be larger or smaller in the full model compared to \texttt{RC+CC}. If the RC electric field is high enough to thermalize the beam in cases with and without runaways, then a larger fraction of the beam is thermalized while neglecting runaways, which can result in a lower potential drop. This is evident through the comparison of the thermalization distance in the 1~MK or 33~MK atmospheres and their corresponding potential drop, for example. With runaways, the potential drop stays low enough to keep all beam electrons from being thermalized in the corona in all cases with a coronal temperature $T>4$~MK. 

The opposite scenario is also possible, i.e., the runaway fraction can be large enough to reduce the total potential drop without thermalizing any fraction of the beam, where warm target effects would dominate in \texttt{RC+CC}. This is the case in the 4~MK and 10~MK atmospheres for fluxes larger than F11 and 4F11, respectively. This non-linearity of the return current losses emphasizes the importance of accounting for the reduction of $\mathscr{E}_{RC}$ by both the thermalization in a warm target and the acceleration of runaways. 

Neglecting the effect of runaway electron acceleration for all the conditions to the left of the $1\%$ runaway threshold is acceptable under the same parameters of cutoff energy, electron spectral index, loop half-length and coronal density.

\subsubsection{Low-energy cutoff, loop-length, spectral index and density dependence}

\begin{figure*}[bth!]
\centering
   \includegraphics[width=0.98\textwidth]{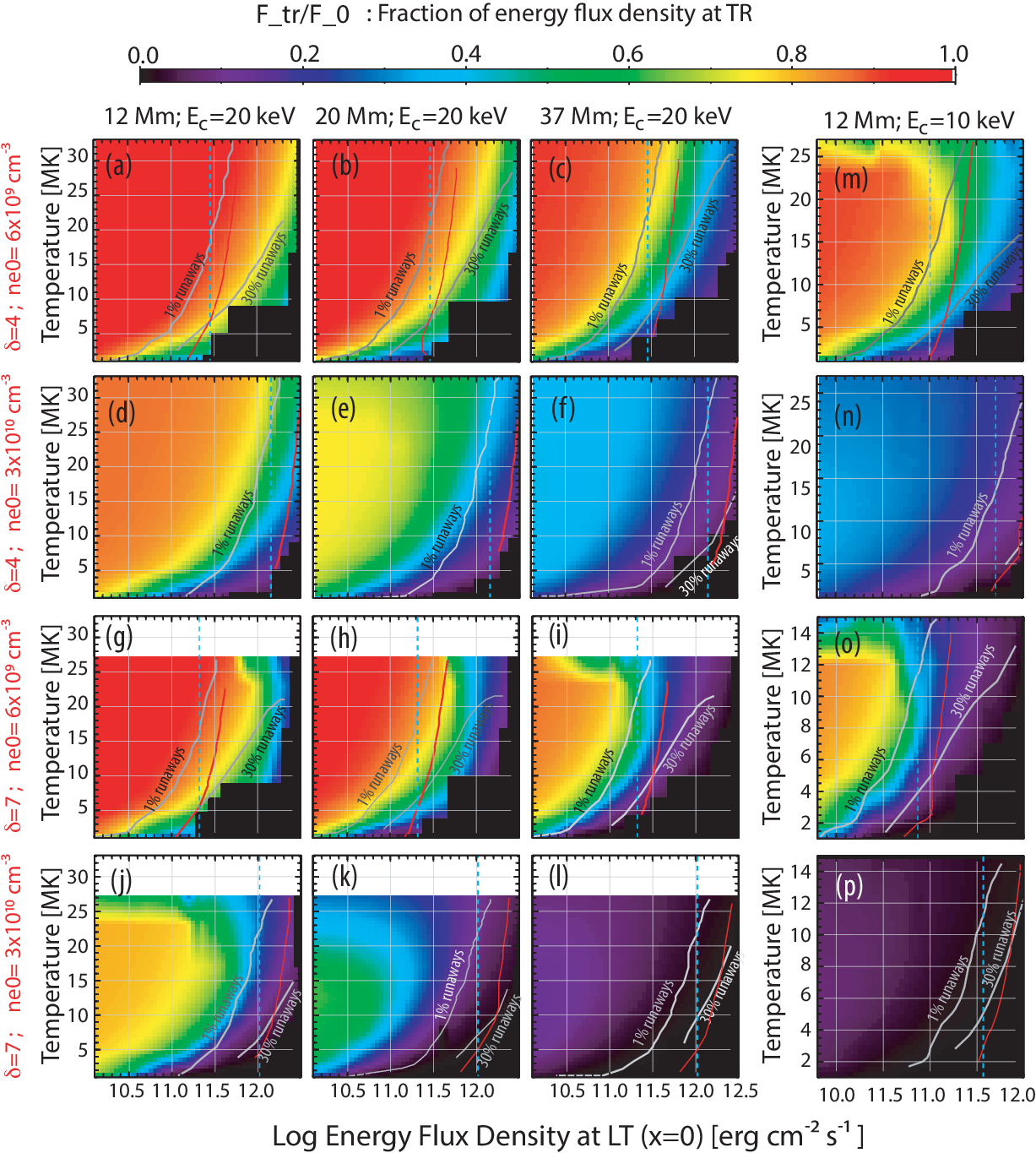}
   \caption{Fraction of energy flux density $\mathscr{F}_{TR}/\mathscr{F}_0$ as a function of injected energy flux density at the looptop $\mathscr{F}_0$ in erg cm$^{-2}$ s$^{-1}$ on the x-axis and the coronal temperature in MK on the y-axis. The red regions are where none of the energy is deposited above the transition region. The coronal (ambient) density and nonthermal electron spectral index are shown on the left of each row. The loop half length is indicated at the top of each column.  The low-energy cutoff for the three left columns and the right column is 20~keV and 10~keV, respectively.  The x- and y-axis are not the same for the 10~keV and 20~keV calculations. No calculations were performed for temperatures higher than 27~MK panels g through l, when the injected distribution has E$_c=20$~keV and $\delta=7$ because the entire beam is thermalized in the corona. The gray curves show the 1$\%$ and 30$\%$ runaway fractions, the cyan dashed line is where the beam density is a tenth of the coronal density $n_b/n_e=0.1$, and the red curves represent $\mathscr{E}_{RC}/\mathscr{E}_D=0.15$. Our model requires the inclusion of additional mechanisms for $\mathscr{E}_{RC}/\mathscr{E}_D\gtrsim0.15$. See the main text. }
  \label{fig:20keV}
\end{figure*}

Figure~\ref{fig:20keV} uses the full model (\texttt{RC+CC+Run}) to compute the fraction of energy flux density which reaches the TR. All the parameters used are indicated on the figure. For example, the low-energy cutoff is 20~keV for all rows in the three left panels, and 10~keV in the right panels. The power-law spectral index of the injected electron distribution and the constant coronal density are indicated on the left of each row. 
Note that the coronal temperature and injected flux density ranges are different for $E_c=10$~keV.

A comparison between calculations using different loop half-lengths, which increase from left to right, shows that the energy flux density deposited in the corona is higher the longer the leg of the loop. This is because a larger potential drop is needed to move particles on longer distances. Compare panels (a-c) or (g-i), where the coronal density of $6\times10^9$ cm$^{-3}$ is too low for Coulomb collisions to affect an electron with initial energy at the low-energy cutoff of 20~keV. 

In addition to the fact that a larger potential drop results from the transport of a beam through a longer loop,  a higher co-spatial density i.e., column density, results in more energy loss by collisions \citep[cf. equation 15 in][]{2021ApJ...917...74A}. This is more easily observed in panels (d-f) and (j-l). Let's take for example, F11.2 and 10~MK where the energy flux density at the TR is 0.85, 0.75 and 0.6 for loop half-lengths 12~Mm, 20~Mm and 37~Mm, respectively. 

A larger fraction of the beam is thermalized for softer electron distributions (compare a and g or m and o for example), and for lower low-energy cutoff values (e.g., compare panels d and n). This shows that warm target effects due to the return current losses are significant and reduce the RC electric field below the thermalization distance because a lower fraction of the injected electron flux density is streaming downward.

These plots are useful to define the thresholds for the importance of the energy flux density deposited in the corona.  However, the details of where the energy is deposited are important for constraining the injected electron distribution from knowledge of the electron distribution at the TR. The detailed calculation is also crucial to determine the thermal response of the plasma to the injected electron beam and the energy distribution of particles returning to the injection site at the looptop.

Note that less than 20$\%$ of the injected beam energy reaches the transition region in panels (l) and (p): [$E_c=20$~keV; $\delta=7$; L=37~Mm; $n_{e0}=3\times10^{10}$~cm$^{-3}$], and [$E_c=10$~keV; $\delta=7$; L=12~Mm; $n_{e0}=3\times10^{10}$~cm$^{-3}$], respectively. Since more than $90\%$ of the energy (and the electron number) flux density is deposited in the corona in these cases, HXR footpoints might not be observable with an instrument like RHESSI \citep{2002SoPh..210....3L}. The combination of imaging and spectroscopy can therefore be sufficient to constrain the injected electron beam and co-spatial plasma in which it propagates.

The curves showing the transition from a purely collisional to a quasi-collisionless return current propagation are indicated by the leftmost solid gray curve ($\sim1\%$ runaways) at the looptop.\footnote{The detailed quantities for each panel are given in the appendix.}  The curve where the runaway flux fraction of the return current is $30\%$ at the LT is the rightmost gray curve. The dashed vertical line shows where the injected beam density is a tenth of the coronal density: $n_{b0}=0.1\,n_e$. Note that the threshold for collisionless propagation of the RC is lower in a lower density plasma. This is due to the fact that the Dreicer field is higher in a denser plasma, $\mathscr{E}_D\propto n/T$. Of course the beam-to-RC density ratio is also higher in a denser plasma, allowing an injected electron beam to propagate in a more collisional regime of the RC than in a less dense plasma. However, that also means that Coulomb collisions are larger in the denser plasma, resulting in a larger energy deposition in the corona.

The red curves show where the maximum RC electric field normalized to the Dreicer field is 0.15 $(max(\mathscr{E}_{RC}/\mathscr{E}_D)=0.15)$. The accuracy of our model decreases significantly above this electric field strength, requiring the inclusion of additional physical mechanisms which are neglected by our model. The justification for this limit comes from our requirement of keeping the local runaway generation within a steady state regime where only the high energy electrons in the tail of the return current distribution will run away.  For the lowest electric fields, the runaway tail is dominated by the highest energy electrons of the RC distribution which are more likely to have a pitch angle along the propagation direction parallel to the guiding magnetic field, and where runaway electrons are dominated by the electrons accelerated by the electric field. These electrons, freely accelerated by the electric field, are called the primary runaway component. Other processes of accelerating runaway electrons include knock-on collisions (avalanche) by the primary runaway electrons and become significant for larger electric fields. This is discussed in section 2.1 of \citet[][]{2021ApJ...917...74A} and was demonstrated in \citet[][]{2014CoPhC.185..847L}. There is a consistency up to electric fields on the order of $0.12\,\mathscr{E}_D$ between the runaway growth rate calculated in \cite{1964PhFl....7..407K} which accounts for primary high energy runaway electrons and the full relativistic time-dependent model of \citet[][]{2014CoPhC.185..847L}.

\section{Impact of return-current electric field mirroring of beam electrons}\label{sec:mirror}

In the 1D model, electrons are not mirrored by the return-current electric field.  All of the downward-streaming energetic electrons are thermalized in either the corona or below.  If the electrons have a distribution of pitch angles relative to the ambient magnetic field, however, some of them may lose their component of energy parallel to the magnetic field without being thermalized.  These electrons are mirrored and accelerated upward toward the top of the flare magnetic loop.  The loss of these electrons decreases the downward number and energy flux of the remaining electrons while the mirrored electrons regain the parallel energy lost in the return-current potential drop and are presumably further energized in the acceleration region at the top of the loop.  
The electron momentum can be divided into two components, one parallel to the magnetic field, $p_\parallel$, taken to be positive in the downward direction, and the other component in the plane perpendicular to the magnetic field, $p_\perp$. If the pitch angle is $\theta$ relative to the magnetic field, then $p_\parallel=p\mu$, with $\mu=cos\,\theta$.
The energy associated with $p_{\parallel}$ is $E_{\parallel}={p_{\parallel}^2}/({2m})=E \mu^2$; and the energy associated with the perpendicular component of the momentum is $E_{\perp}=E(1-\mu^2)$, with $E=E_{\parallel}+E_{\perp}$. In the 1D model all electrons have $0^{\circ}$ pitch angle so that $\mu=1$, and $p_{\parallel}=p$, and $E_{\parallel}=E$.

In the presence of the return-current electric field, an electron's pitch angle becomes $90^{\circ}$ where the potential drop $V=E_{\parallel}$, and $E_{\parallel}$ is the value of the electron's parallel energy at injection. Therefore, electrons for which $V>E_{\parallel}$, i.e., those with $\mu<(V/E)^{1/2}$, are mirrored.  Accounting for the model approximation that electrons are thermalized at $E-V=E_{th}$, only particles for which $E>V+E_{th}$ are mirrored. Therefore, only electrons with an injected pitch-angle cosine $\mu<(V/(V+E_{th}))^{1/2}$ are mirrored. 

In general, a larger fraction of electrons can be mirrored in a lower temperature plasma than in a higher temperature plasma. In the lower temperature plasma a larger fraction of the electrons are mirrored before being thermalized ($(V/(V+E_{th}))^{1/2}$ is larger because $E_{th}$ is smaller and $V$ is larger). The highest pitch angle (small $\mu$) electrons are mirrored near the top of the loop, while the smallest pitch angle electrons are mirrored low in the loop.

\begin{figure}[bth!]
\centering
    \includegraphics[width=0.495\textwidth]{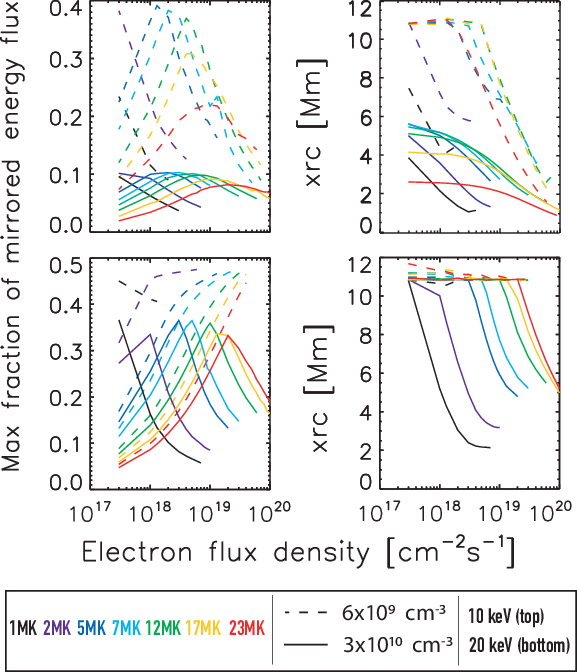}
   \caption{{\textit{Left:}} Maximum fraction of the injected energy flux which could be mirrored by the RC electric field in an initially isotropic distribution.{\textit{Right:}} Thermalization distance in a 12Mm half loop. The top and bottom panels represent cutoff energies of 10~keV and 20~keV.}
  \label{fig:mirror}
  \end{figure}

Here we consider an injected pitch angle distribution that is isotropic from $0^{\circ}$ to $90^{\circ}$ for all electron energies. Integrating the isotropic distribution over pitch angle, from $\mu=0$ to $\mu=(V/E)^{1/2}$, shows that the energy dependence of the mirrored electron distribution is that of the original downward distribution times $\sqrt{V/E}$ i.e., \begin{equation} \mathscr{F}_{iso}= \int_{E_c(TR)}^\infty E dE\, F_{1D}(x_{TR},E)\sqrt{\frac{V_{TR}}{E}},\end{equation} where $\mathscr{F}_{iso}$ is the maximum energy flux density mirrored upward by the RC electric field if the electron distribution at injection at the looptop is isotropic. It is steeper by $E^{-1/2}$ than the downward distribution.
The downward energy flux of electrons is obtained by first integrating over pitch angle from $\mu=(V/E)^{1/2}$ to $\mu=1$. This gives $1-(V/E)^{1/2}$ in the integrand. The downward energy flux is therefore the downward energy flux without mirroring minus the energy flux of the mirrored electrons. Since the lowest energy electrons are preferentially mirrored, the loss of downward flux is greater for injected energy spectra that are steeper than for flatter spectra.

Figure~\ref{fig:mirror} shows the maximum mirrored energy flux density if the injected electron distribution is isotropic instead of being initially fully beamed. The spectral index is 4 and loop half-length is 12 Mm. The higher the injected electron flux density, the larger the RC electric field and therefore the larger the reduction of the parallel component by the electric field. This trend reverses if a fraction of the electron beam is thermalized, and if the thermalization happens closer to the looptop, i.e., a larger fraction of the beam is thermalized and therefore the mirroring is lower.

These calculations are useful for determining when electron mirroring is likely to significantly reduce the downward electron energy flux. Since the electrons are mirrored along the length of the flare magnetic loop, mirroring reduces the downward number flux and therefore the return-current electric field strength and potential drop along the length of the loop. For accurate numerical results, the mirroring must be included in the calculation of these quantities. Further discussion can be found in Section 5.1 in \citep{2021ApJ...917...74A}, which has an estimate of the reversed nonthermal electrons using \citet{2020ApJ...902...16A} compared to the nonthermal electrons accelerated out of the thermal plasma (runaways). These two effects, beam reversed electrons and acceleration of runaways, need to be studied systematically for a wide range of initial pitch angle distributions. 

\section{Return current stability}\label{sec:instability}
The solar corona, where the primary acceleration of electrons is thought to occur, is a quasi-collisional/collisionless plasma depending on the specific conditions for each flare.

Our model accounts for binary collisions between electrons and ions. However, if the RC drift speed $v_d$ is large enough to reach the threshold for generating current-driven instabilities, then a model which accounts for them needs to be developed. Specifically, the collision frequency may be enhanced as a result of the passage of the return current in an unstable plasma \citep[e.g.,][]{2005SSRv..121..237B}. Since both the resistivity and the Dreicer field are directly proportional to the collision frequency, it is necessary to figure out whether the resistivity and the Dreicer field might be enhanced due to a RC-driven instability compared to the Spitzer values. This will significantly modify the fraction of accelerated runaway electrons, and therefore the amount of energy deposition in the solar atmosphere.

Although our model does not account self-consistently for the eventual generation of return-current-driven instabilities, we show in this section that as the injected flux density is increased, the threshold for acceleration of runaway electrons is lower than that to drive a current instability if the electron-to-ion temperature ratio is sufficiently small ($T_e/T_i\lesssim4$). We use the current instability thresholds from \cite{1982PhFl...25.1183M,1985ApJ...293..584H}, where the lowest threshold is the electrostatic ion cyclotron which is  $v_d\sim0.4\,v_{Te}$, where $v_{Te}=\sqrt{2k_BT_e/m_e}$, $T_e=T_i$, and decreases to $v_d\sim 0.14 v_{Te}$ at $T_e\sim 4\,T_i$. Note that the thresholds from these authors were adjusted to be normalized to the same electron thermal speed definition we use in this paper and \cite{2021ApJ...917...74A}. Additionally, the injected flux density threshold which drives a drift speed large enough for generating these instabilities is increased when runaway electrons are taken into account.

\begin{figure}[bth!]
\centering
 \includegraphics[width=0.49\textwidth]{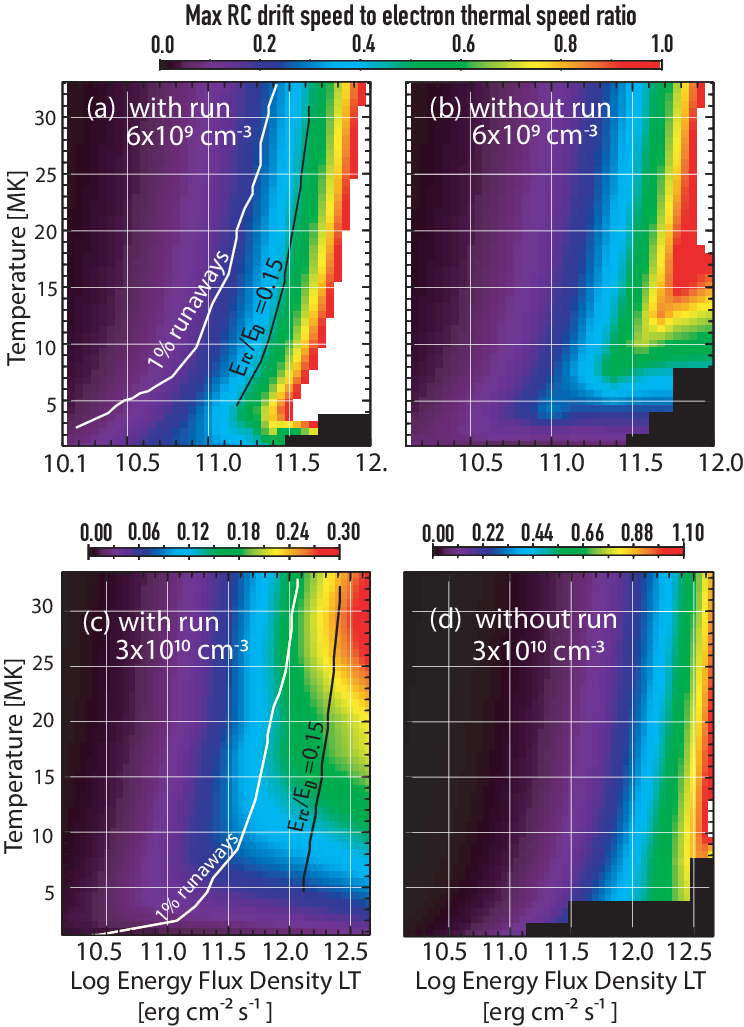}
   \caption{Maximum return current drift speed $v_d$ to electron thermal speed $v_{Te}$ ratio. We use $\delta=4$, E$_c$=20~keV and loop half length of 20~Mm. }
  \label{fig:vd}
\end{figure}

Figure~\ref{fig:vd} shows the maximum fraction of the RC drift speed and electron thermal speed in models with (\texttt{RC+CC+Run}) and without (\texttt{RC+CC}) inclusion of runaway electrons as a function of the log of the injected nonthermal flux density (x-axis) and the coronal temperature (y-axis). The top two and bottom two panels use a coronal co-spatial density of $6\times10^9$ cm$^{-3}$ and $3\times10^{10}$ cm$^{-3}$, respectively. The maximum drift speed is reached anywhere along the loop legs and corresponds to its value at the thermalization distance (see, for example, the magnitude of the heating rate at the thermalization distance $X_{RC}$, i.e., the dashed lines in Figure~\ref{fig:heatrate}). Note that the maximum normalized RC drift speed is comparable in the low coronal density case (top panels), but this speed is much lower in the higher coronal density case (bottom panels). In the lower co-spatial density case, the RC electric field is sufficiently large to start electron thermalization at different heights resulting in very different heating profiles but comparable maximum drift velocity. Since we are interested in whether the drift speed ever reaches an instability threshold, the maximum value is sufficient. This is similar to the maximum heating rate from the detailed example in Figure~\ref{fig:heatrate}.

Using the thresholds of \cite{1982PhFl...25.1183M}, first note in Figure~\ref{fig:vd} that the electron flux density threshold for generating runaway electrons, say where runaway electrons are 1\% of the returning flux at the looptop (solid white curve on the left panels), is lower than the instability threshold at the normalized velocity drift fraction of 0.4 (cyan in panel a and outside the plotted range in panel c). This indicates that runaway electron acceleration takes precedence over instability generation. If however, the electron to ion temperature ratio is larger than 4, the drift velocity threshold drops to $0.14\,v_{Te}$, then whether runaways would be accelerated or a current-driven instability depends on the coronal density and temperatures as well as the injected electron fluxes. If a current instability enhances the effective collision frequency, then we expect the resistivity and the effective Dreicer field to be enhanced, and therefore the fraction of runaways to be reduced. 

These two collisionless regimes, where runaway electrons or current instabilities dominate the dynamics of transport, are energetically very different. In the first case of runaway acceleration, the RC electric field is expected to get reduced compared to the collisional Spitzer case, thereby heating the plasma to a lower degree. In the second case of RC driven instabilities, the effective resistivity is enhanced, resulting in increased Joule heating.

This work shows the conditions for binary collisions to dominate the beam/RC system dynamics, and where collisionless effects including runaway acceleration or a current-driven instability might need to be taken into account self-consistently. Runaway electrons become significant in modifying the energy of the beam/RC system for lower injected fluxes if the electron and ion temperatures are equal.  A self-consistent treatment of collisional and collisionless effects, which includes both the acceleration of runaway electrons and wave-particle interactions is warranted at $T_e\gtrsim 4\,T_i$ and sufficiently low densities.

\section{Inspection of previously published solar flare analyses}
\label{sec:obs}

Now that we have systematically computed the reduction of the nonthermal energy flux density due to the return current and Coulomb collisions, we demonstrate in this section that several flares, previously interpreted using the collisional thick target model (CTTM), need to be re-analyzed to account for energy loss mechanisms not included in the CTTM, i.e., RC losses with the eventual need to include the effect of runaway electrons in a warm-target. We first compile the nonthermal beam parameters from various published papers and then discuss the conditions under which there is an inconsistency between the CTTM and the fact that return currents are negligible. Since this also depends on the co-spatial parameters of coronal temperature and density and importantly, the cross-section area of the loop, we discuss the possibilities as a function of these parameters.
Two notable examples, with reported  nonthermal energy flux densities at or above F12 are \citet{2003ApJ...595L.111W,2011ApJ...739...96K}, for SOL2002-Jul-23T00:25 and SOL2006-Dec-06T18:50, respectively. 

In \citet[][hereafter, W03]{2003ApJ...595L.111W}, the nonthermal beam density above 20~keV is $n_b=10^{10}$ cm$^{-3}$, which is equivalent to 2F12 for a low-energy cutoff of 20~keV and spectral index of $\delta=4$, according to equation~\ref{eq:eflux} and $F_{e0}=n_{b}\, v_{b0}$ where $n_{b}$ is the injected electron density and $v_{b0}$ the average beam velocity at the looptop, $v_{b0}=(\delta-1)/(\delta-1.5)\,(2\,E_c/m)^{1/2}$. The authors also deduce from seemingly co-spatial TRACE 195\AA\, images, the co-spatial coronal temperature of the beam to be below 20~MK, so we will use that as an upper limit to the co-spatial coronal temperature, although this does not affect the conclusions.

In \citet[][hereafter, K11]{2011ApJ...739...96K}, the nonthermal energy flux density above 18~keV is 5F12, which was calculated by dividing the electron flux density, deduced from fitting the HXR peak spectrum using the CTTM, by the area of the footpoints, deduced from the Solar Optical Telescope onboard Hinode.

The loop half-lengths from both flares are below 12~Mm \citep[Table~3 in][]{2017ApJ...851...78A, 2003ApJ...595L.119E}, the spectral indices are between 4 and 4.5, and the low-energy cutoff values used are 18-20~keV. We therefore perform the corresponding calculations using our full model \texttt{RC+CC+Run} and three coronal densities.

\begin{figure}[bth!]
\centering
  \includegraphics[width=0.495\textwidth]{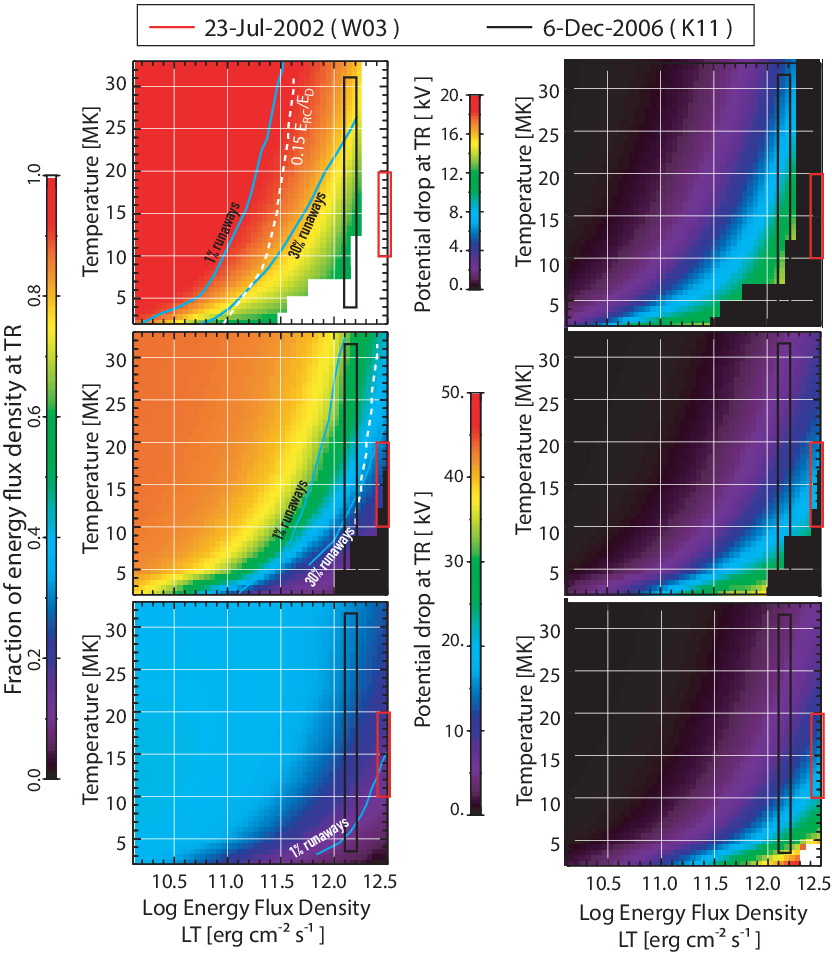}
   \caption{\textit{Left:} Fraction of energy flux density reaching the TR. \textit{Right:} RC potential drop at the TR in kV. From top to bottom: Calculations using a co-spatial coronal density of $6\times10^9$, $3\times10^{10}$, and $1\times10^{11}$ cm$^{-3}$, respectively. The other parameters used are a loop half-length of $\sim$ 12~Mm, a low-energy cutoff of 20~keV, and spectral index $\delta=4$. The red and black rectangles represent the deduced energy flux densities using the CTTM from W03 and K11, respectively.}
  \label{fig:flares}
\end{figure}

Figure~\ref{fig:flares} shows the deduced energy flux density over-plotted on the modeled fraction of energy flux density which reaches the TR. The curves showing the 1$\%$ and 30$\%$ fractions of the return current carried by runaway electrons at the looptop are represented in blue. Also plotted in white is the curve where the accuracy of our model reduces significantly around $\mathscr{E}_{RC}=0.15\mathscr{E}_D$. 

Clearly, no matter what co-spatial parameters are used, the deduced energy flux densities in these two flares using the CTTM are too high to neglect the return current losses, as shown in the right panels of figure~\ref{fig:flares}. Additionally, if the co-spatial density is $1\times10^{11}$ cm$^{-3}$, the fraction of the RC carried by runaway electrons is negligible for both flares. If the co-spatial density is $3\times10^{10}$ cm$^{-3}$, the flare in W03 requires the inclusion of both the RC losses and the effect of runaway electrons, and this fraction of runaways decreases with increasing coronal temperatures. For the same coronal density, the flare in K11 is outside of the range where our model is accurate, and collisionless effects as discussed in section~\ref{sec:instability} will dominate the dynamics of the beam/RC system propagation. Finally, for a density $6\times10^9$ cm$^{-3}$, the need to develop a collisionless model of the beam/RC propagation is increased, however, note that the density of the background is lower than the beam density.

These results show that these flares need to be re-analyzed using a model which accounts for the return current electric field deceleration of the nonthermal beam distribution.

The flare K11 was also analyzed using the return current collisional thick-target model of \citet{2012ApJ...745...52H} in AH17. The authors found that the spectral breaks in several time intervals during the HXR peak time can be explained by a RC potential drop, even when using a footpoint area deduced from HXR observations. This area might be unresolved and can be up to an order of magnitude higher than that deduced from white-light observations \citep[][]{2014ApJ...793...70M,2011ApJ...739...96K}. Observations of newly activated sources indicating where energy is being deposited can be as small as 100~km in H$_{\alpha}$ \citep[][]{2014ApJ...788L..18S}.

The W03 flare was omitted from the analysis of AH17 because of the criterion to only include spectral breaks that cannot be explained by non-uniform ionization or albedo. The possibility to explain the spectral breaks in SOL2002-Jul-23 by non-uniform target ionization was studied by \citet[][]{2003ApJ...595L.123K}. However, this does not rule out the contribution of the return current to the energy losses because an injected energy flux of F12 is inconsistent with negligible effects of the RC on energy deposition.

A refined analysis of these flares should be performed. This should include RC losses along the legs of the loop, the shape of the HXR spectra as a function of time and the local and global energetics of the flare.

\section{Discussion}
\label{sec:discussion}

\subsection{Understanding the beam/RC system dynamics}
To understand the electron beam/return-current system dynamics, it is useful to think about the hierarchy of importance of beam and ambient plasma parameters on the return current losses.

For example, at the lowest injected flux densities, the return current is governed by Ohm's law where the RC electric field is proportional to the beam flux density and the resistivity, which, in the simplest case of binary collisions between electrons and ions, is proportional to $T_e^{-3/2}$. This means that if the flux density is low enough (say $10^{15}$ electrons cm$^{-2}$ s$^{-1}$), or the coronal temperature is high enough (say 40 MK), then the other parameters of co-spatial density, nonthermal electron energy distribution spectral index, low-energy cutoff, and loop length do not matter for energy losses in solar flares. RC losses are negligible under the above-mentioned conditions. 

If, however, the flux density and temperature are high and low enough respectively, then all beams with the same total number flux density (for example $F_{e0}=10^{18}$ cm$^{-2}$ s$^{-1}$ equivalent to 2.4F10 and 4.8F10 if $\delta=4$ and the low-energy cutoffs are 10 and 20~keV, respectively), and co-spatial temperatures will experience the same local RC losses, down to the distance at which the beam starts being thermalized. The reduction of the energy flux density below the thermalization distance is then higher for a softer electron distribution (lower $\delta$) and a lower low-energy cutoff. The beam with a lower low-energy cutoff is thermalized higher in a loop both because a lower RC potential drop is required to thermalize such a beam and because Coulomb collisional losses are larger for a lower energy electron.  Therefore, in a return current model with a cold target assumption, where electrons are not allowed to get thermalized and lost from the injected electron flux if their energy becomes comparable to the plasma in which they propagate, the RC electric field and the total potential drop are overestimated below the thermalization distance.

The same potential drop is expected for an injected electron beam streaming in plasmas with different densities unless (1) the beam starts to be thermalized through Coulomb collisions or (2) if runaway electrons become significant.  In the first case, of course, the larger CC losses are experienced by lower energy electrons, and by beams streaming through a larger column density, so below the thermalization distance, the flux density is reduced and so is the RC electric field. In the second case, the dependence on the density as larger fractions of runaway electrons are accelerated can be noted by comparing the results in Figure~\ref{fig:flares}, for example, where the potential drop is the same in atmospheres with different densities but the fraction of runaways is higher in a less dense atmosphere. This is because the Dreicer field is proportional to the density, i.e., a smaller fraction of runaways is accelerated in a denser plasma. 

The loop length is possibly the easiest parameter to constrain from observations. It is easy to see that the larger the distance the electrons must travel, the larger the total potential drop. However, the nonthermal electron flux thresholds for accumulating $1\%$ or a higher fraction like $30\%$ runaway electrons at the looptop are independent of the loop length. However, the energy deposition above the transition region is larger in a longer loop due to RC losses and Coulomb collisions.

\subsection{Thermal response}
It is important to note that the initial conditions in hydrodynamic codes used for estimating the thermal response of a plasma to the injection of a nonthermal electron beam are highly sensitive to the behavior of the return current dynamics. Significantly different spatial and temporal evolutions are expected if the initial conditions are from pre-flare conditions in radiative equilibrium (e.g., VAL-C atmospheres \citep[][]{1981ApJS...45..635V} with apex temperature at 1-5~MK and apex density lower than $10^{10}$ cm$^{-3}$), as it it usually assumed \citep[e.g.,][]{2020ApJ...895....6G,2015ApJ...813..133R,2022ApJ...928..190K}, or if the coronal plasma is already at the highest temperature and density observed. Also relevant as the energy injection propagates to different flux tubes or magnetic threads, are the initial co-spatial temperature and density profiles. For example, in \citet[][]{2015ApJ...813..133R}, separate threads were defined based on the "burstiness" of the HXR lightcurve, and the initial conditions of each burst are similar to quiet sun conditions, since this does not significantly change the dynamics of where and how much energy is deposited along the loop legs, if return currents are not taken into account. While this is acceptable for the lowest injected flux densities, it is increasingly important to properly account for return current effects, including its appropriate regime of propagation. This dependence on the co-spatial plasma parameters can help constrain the instantaneous temperature and density of the co-spatial plasma to the nonthermal electron beam. 
A new grid of simulations similar to the widely used and pre-calculated simulations F-CHROMA\footnote{Simulations can be accessed through \url{https://star.pst.qub.ac.uk/wiki/public/solarmodels/start.html}} 
 but with the inclusion of return current losses will be the subject of future work.

\subsection{Neglected mechanisms}
\begin{enumerate}
\item \textbf{Co-spatial ion beams:} The accelerated ion distribution if it is co-spatial to the electron beam should reduce the RC as derived in this paper. Although there is evidence for the ion footpoints being significantly separated from the electron footpoints \citep[][]{2003ApJ...595L..77H,2006ApJ...644L..93H,2004ApJ...602L..69E}, these ions have energies higher than $\sim 1$ MeV, and could be a different population of ions compared to an unobserved lower energy ion distribution of energies. Some PIC+hydrodynamic kglobal simulations (Z. Yin, J. F. Drake, and M. Swisdak, submitted) indicate that the energy of accelerated ions is of the same order of magnitude as the electrons. If indeed, these ions are always accelerated when electrons are, and if they propagate co-spatially to the electrons, then they would reduce the return current electric field as calculated in our model. The extent of this reduction has not been investigated yet, to the best of our knowledge.
\item \textbf{Loop cross section expansion:} If the area of the loop expands upward, then the injected electron flux density is reduced by the same factor at the looptop resulting in a reduction of the magnitude of the local RC electric field. \citet[][]{2022ApJ...933..106R,2024ApJ...967...53R} found that the heating in expanding loops affects the temporal evolution by lengthening the cooling times for example.

\item \textbf{Pitch-angle scattering of the nonthermal beam electrons:} Even without return current losses, pitch angle scattering can result in a lower energy deposition in the corona \citep[cf. dashed curves in][section 5 and figure 8]{2021ApJ...917...74A}. \citep[][]{2017ApJ...835..262B,2019ApJ...880..136J}. 
\item \textbf{RC electric field mirroring:} If the initial pitch-angle distribution is isotropic then the energy flux density can be reduced by as much as $50\%$ compared to the fully beamed injection, as calculated in Section~\ref{sec:mirror}. A model which accounts for both the pitch-angle distribution of the nonthermal beam and runaway electrons is needed to properly account for these cases.

\item \textbf{Turbulence due to magnetic field fluctuations:} If turbulence due to magnetic field fluctuations increases the electrical and thermal resistivities \citep[][]{2016ApJ...833...76B,2018ApJ...865...67E} then the thresholds for accelerating runaway electrons is reduced, because the anomalous effective Dreicer field is increased as it is proportional to the effective collision frequency. This results in an increased local RC electric field.

\item \textbf{Magnetic field mirroring:} Magnetic mirroring could also decrease the downward propagating electron beam resulting in a reduction of the magnitude of the RC electric field. Although \citet[][]{1995A&A...304..284Z} report that the magnetic field convergence does not significantly change the energy losses but reduces the angular scattering of electrons along the loop legs. Their model includes Ohmic RC losses, CCs, pitch-angle scattering and magnetic convergence in a cold target. Since an isotropic initial pitch-angle distribution also reduces the downward propagating beam flux density, the coupled effect of RC electric field mirroring, pitch-angle scattering and magnetic convergence needs to be systematically studied. 

\item \textbf{Time scales:} We assume that the injection time is long compared to the time for the RC to reach a steady-state. This happens on collisional time scales (cf. equation 4 in A21). The longest collision time for parameters in this paper is 0.42~s for $T_e=33$~MK and the lowest density of $6\times10^9$ cm$^{-3}$. 

\end{enumerate}

\section{Summary of main results}
Deceleration by the co-spatial return current is a non-negligible effect for a wide range of solar flare parameters. The extent of the contribution of the RC to the deceleration of the beam and energy deposition along the loop is highly sensitive not only to the injected nonthermal beam electron energy distribution but also to the co-spatial temperature and density profiles.

Higher potential drops can be expected in collisional plasmas, i.e., higher density plasmas, because fewer runaways and larger injected fluxes can stably stream along the loop legs. 

Although comparable total potential drops at the transition region may be deduced with models that include or neglect the effect of runaway electrons, the magnitude of the energy deposition in the corona is different. Twice as much energy is deposited at the looptop when runaways are neglected in the example in section~\ref{sec:example}, and a factor of 4 less energy is deposited around 9~Mm because of thermalization of lower energy electrons. 

Runaway electrons return to the looptop but it is unclear whether they become part of the accelerated (or at least injected) electron distribution or stream along the other loop leg. 

A number of flares need to be re-analyzed to account for return current losses. This includes microflares through the largest flares. However, in microflares, the largest energy deposition by RC losses is expected closer to the looptop. This is due to the lower values of the low-energy cutoff which are easier to thermalize.

The initial conditions in simulations of the thermal response need to be reviewed, and/or return currents systematically included. The need to also include the dynamics of runaway electrons are increased for larger flux densities and lower co-spatial temperatures. This paper provides these conditions for solar and stellar flare conditions.

A self-consistent treatment of collisional and collisionless effects, which includes both the acceleration of runaway electrons and wave-particle interactions is warranted at $T_e\gtrsim 4\,T_i$ and sufficiently low densities. For sufficiently low $T_e$ to $T_i$ ratios, the electron nonthermal flux requirement to accelerate runaway electrons is lower than that to drive a current instability.

\section*{Acknowledgements}

This work was supported through the following NASA grants  80NSSC23K0043 and 80NSSC23K0448 to the University of Maryland. M.A. also acknowledges interactions with J. Allred, G. Emslie, and J. Reep. who are collaborators under the second grant above; and E. Kontar and J. Drake for useful discussions.

\bibliography{main.bib}

\section*{Appendix}
The calculation of the energy flux density reduction at the transition region, fraction of runaway electrons at the looptop, the maximum normalized RC electric field magnitude (at the thermalization distance $x_{RC}$), the potential drop at the transition region, the maximum RC drift speed normalized by the electron thermal speed, and the thermalization distance in Mm.

\begin{figure*}[bth!]
\centering
  \includegraphics[width=0.75\textwidth]{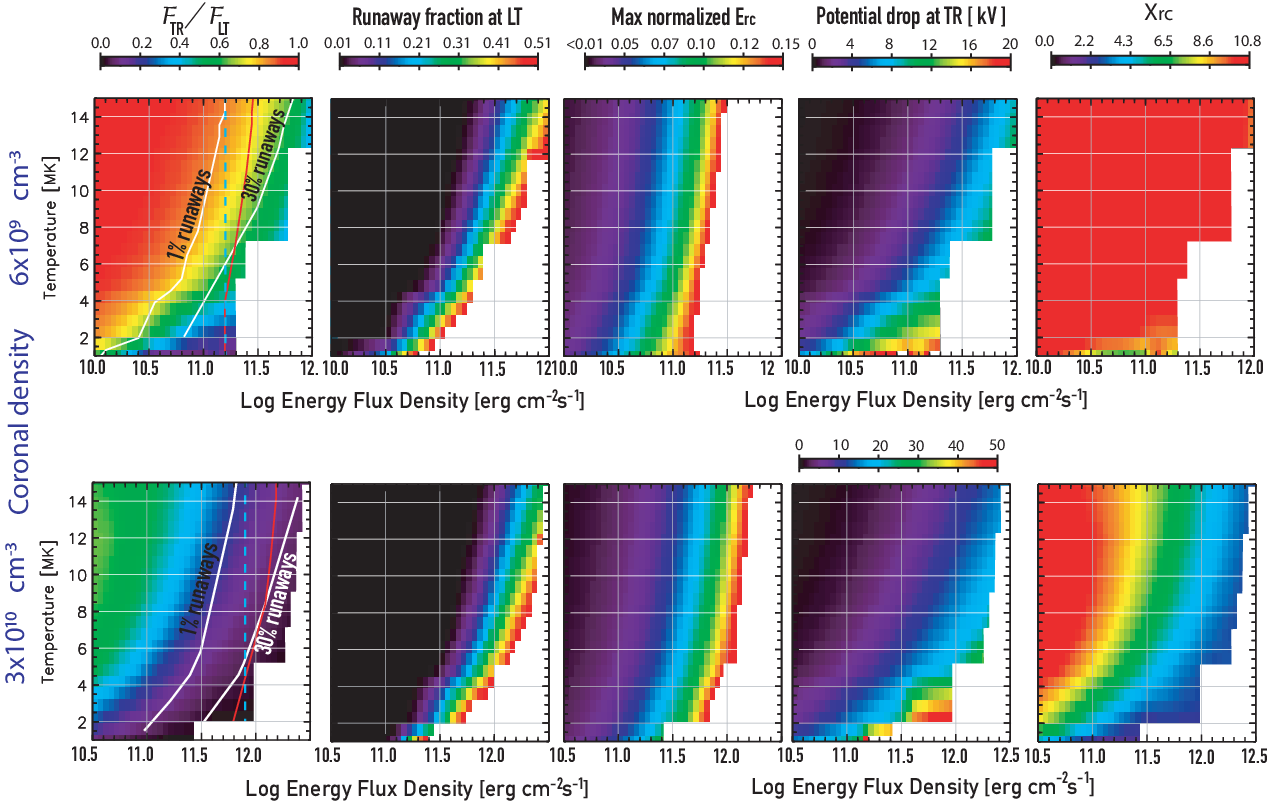}
   \caption{All calculations use 12~Mm, 15~keV, $\delta=5$. The coronal density in the top and bottom panels are $6\times10^9$ cm$^{-3}$ and $3\times10^{10}$ cm$^{-3}$, respectively. From left to right: Fraction of energy flux density deposited above the transition region as a function of injected flux density (x-axis) and coronal temperature (y-axis), runaway fraction at the looptop, the maximum normalized (to the Dreicer field) RC electric field, the total potential drop at the transition region, and the thermalization distance from the looptop. Note that the maximum distance plotted is at the transition region. Note the different ranges between top and bottom panels: The x-axis covers 10 to 12 and 10.5 to 12.5, and the colorbars are the same except for the potential drop at the transition region.}
  \label{fig:15keV}
\end{figure*} 

\begin{figure*}[bth!]
\centering
  \includegraphics[width=0.95\textwidth]{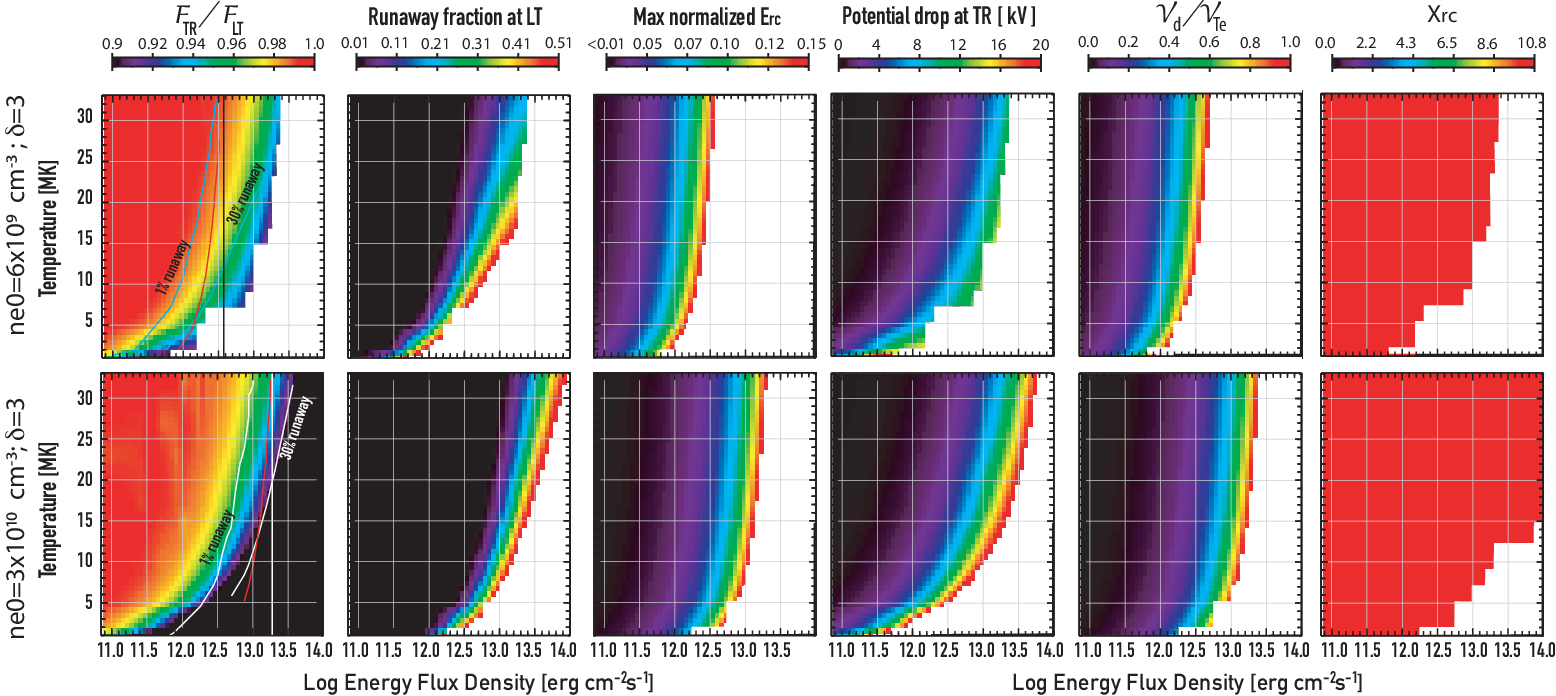}
   \caption{Same as figure~\ref{fig:15keV} with 12~Mm, 85~keV, $\delta=3$. Note the different colorbar range in the fraction of energy flux density deposited above the transition region. }
  \label{fig:85keV}
\end{figure*} 

\begin{figure*}[bth!]
\centering
  \includegraphics[width=0.85\textwidth]{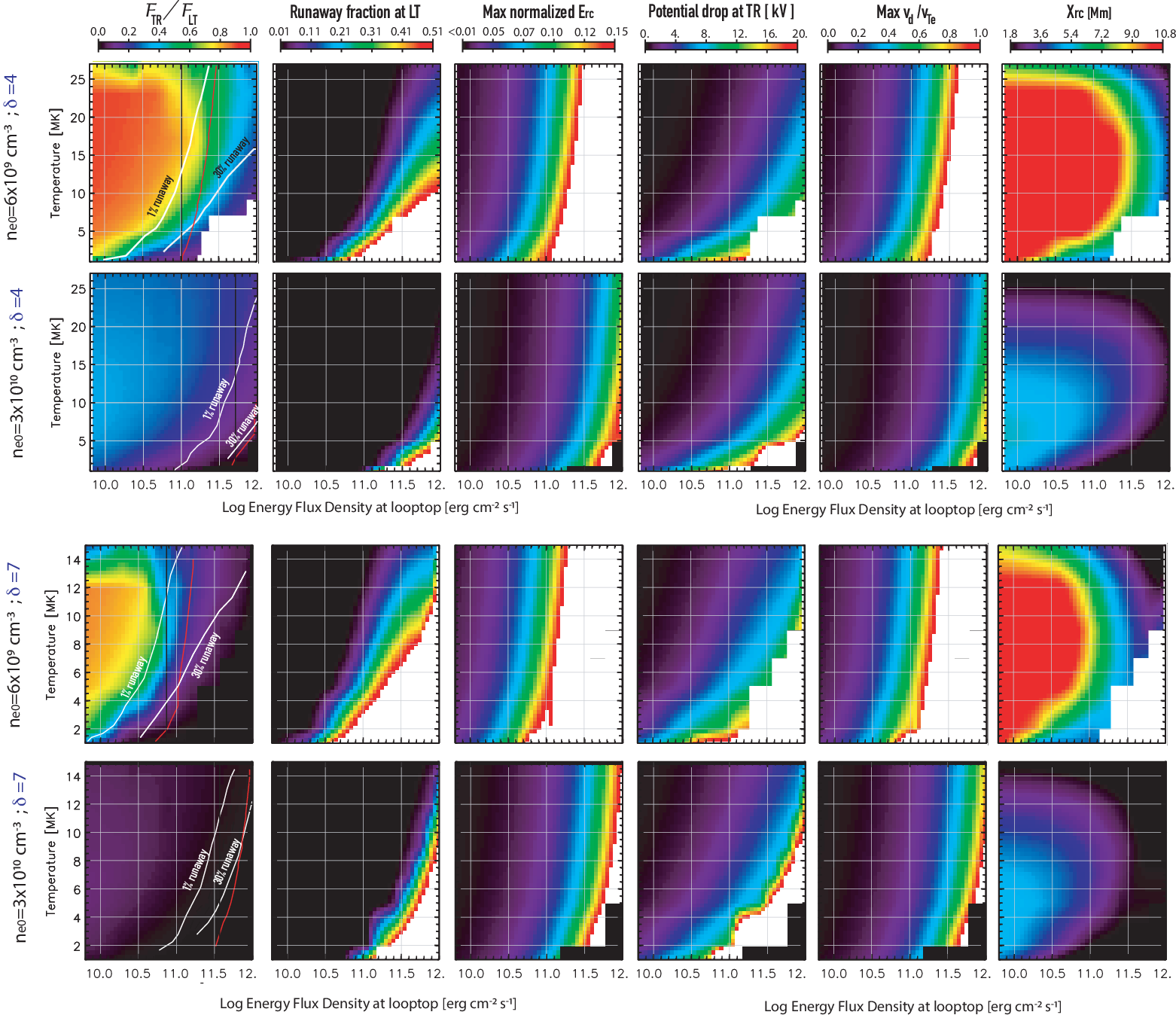}
   \caption{Same as Figures~\ref{fig:85keV} and \ref{fig:15keV}. Parameters used are a loop half-length of 12~Mm, $E_c=$10~keV. Note the different temperature ranges in the top two and bottom two panels.}
  \label{fig:10keV_all}
\end{figure*} 

\begin{figure*}[bth!]
\centering
  \includegraphics[width=0.95\textwidth]{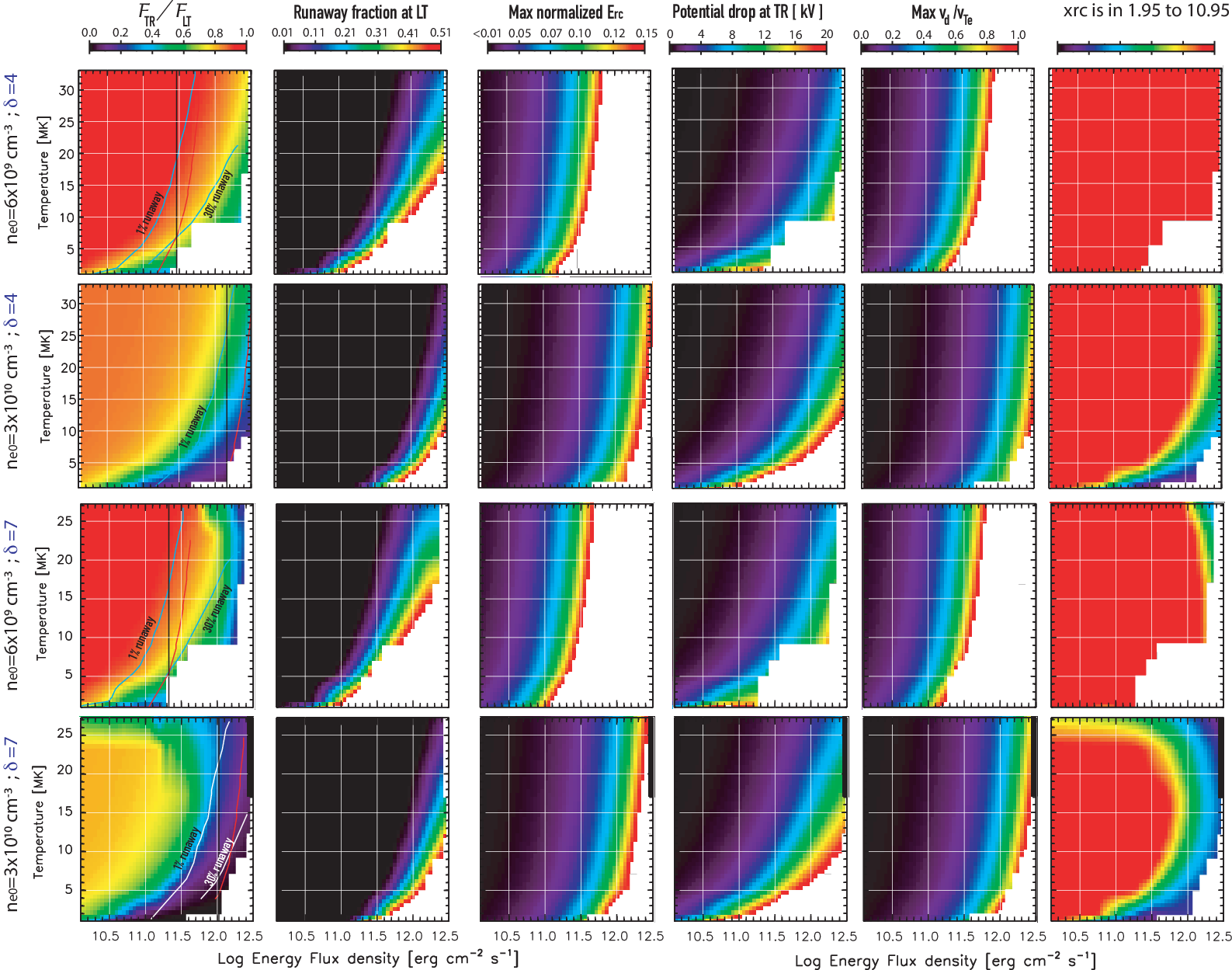}
   \caption{Same as Figure~\ref{fig:10keV_all} with a loop helf-length of 12~Mm and low-energy cutoff of 20~keV}
  \label{fig:20keV_all}
\end{figure*} 

\end{document}